\documentstyle[osa,manuscript,graphicx]{revtex}

\def\r2{\sqrt 2}
\def\beq{\begin{equation}}
\def\eeq{\end{equation}}
\def\beqn{\begin{eqnarray}}
\def\eeqn{\end{eqnarray}}

\def\sinW2{\sin^2\theta_W}

\def\mz2{M_{z}^2}
\def\c2b{\cos 2\beta}

\def\mz{M_z}

\def\Fq2{F_{2}(q^2)}

\def\beq{\begin{equation}}
\def\eeq{\end{equation}}

\def\gmin2{(g-2)_\mu}
\def\ecoup{{e \over {{\sqrt2}\sin{\theta_W}}}}
\def\winomass{m_{\tilde W_a}}

\def\sec2w{sec^2\theta_W}

\def\r2{\sqrt 2}
\def\beq{\begin{equation}}
\def\eeq{\end{equation}}
\def\beqn{\begin{eqnarray}}
\def\eeqn{\end{eqnarray}}

\def\sinW2{\sin^2\theta_W}

\def\mz2{M_{z}^2}
\def\c2b{\cos 2\beta}

\def\mz{M_z}

\def\Fq2{F_{2}(q^2)}
\def\sq2{\sqrt{2}}

\def\sec2w{sec^2\theta_W}

\begin{document}
\begin{flushright}
{CERN-TH/2001-124 }\\
\end{flushright}
%
\begin{center}{\Large \bf PROBE OF SUSY AND EXTRA DIMENSIONS BY 
THE BROOKHAVEN g-2 EXPERIMENT}\\
\vskip.25in
{Pran Nath }

{\it
 Theoretical Physics Division, CERN CH 1211, Geneva, Switzerland} 
\\
\it Department of Physics, Northeastern University, Boston, 
MA 02115-5000, USA\footnote{Permanent address}\\%

\end{center}

\begin{abstract}                
A brief review is given of $a_{\mu}=(g_{\mu}-2)/2$ as a probe of 
supersymmetry and of extra dimensions. It is known since 
the early to mid nineteen
eightees that the supersymmetric electro-weak correction to $a_{\mu}$
 can be as 
 large or larger than the Standard Model electro-weak correction
 and thus any experiment that proposes to test the Standard Model electro-weak
 correction will also test the supersymmetric correction 
  and constrain
 supersymmetric models. 
 The new physics effect seen in the Brookhaven (BNL) experiment
 is consistent with these early expectations. Detailed analyses 
 within the well motivated supergravity unified model show that
 the size of  the observed difference ($a_{\mu}^{exp}-a_{\mu}^{SM}$)
 seen at Brookhaven implies upper limits on sparticle masses in a 
 mass range accessible to the direct observation of these particles
 at the Large Hadron Collider. Further, analyses also show that the
 BNL data is favorable for the detection of supersymmeteric dark 
 matter in direct dark matter searches. The effect of large extra 
 dimensions on  $a_{\mu}$ is also discussed. It is shown that with
 the current limits on the size of extra dimensions, which imply
 that the inverse size of such dimensions lies in the TeV region,
 their effects on $a_{\mu}$ relative to the supersymmetric contribution
 is small and thus  extra dimensions do not 
 produce a serious background to the supersymmetric contribution.
It is concluded that the analysis of the additional data currently 
underway at Brookhaven as well as a reduction 
of the hadronic error will help pin down the scale of weak scale 
supersymmetry even more precisely.
\end{abstract}

\section{ Introduction}
The  topics we discuss in this paper consist of the supersymmetric
  electro-weak effect  on $a_{\mu}$,  and the
 implications of the precise Brookhaven (BNL) $a_{\mu}$ result for the
 direct detection of supersymmetry at accelerators and 
 in the direct search for dark matter. We will also discuss the effect
 of extra dimensions on $a_{\mu}$.
We  begin by reviewing the situation with regard to the Standard
Model contribution
which consists of 
$a_{\mu}^{SM}=a_{\mu}^{qed}+a_{\mu}^{had} + a_{\mu}^{EW}$.
The qed corrections have been computed to $O(\alpha^5)$
(for a review see Refs.\cite{czar}), 
 the Standard Model electro-weak correction
is computed to one loop\cite{fuji} and two loop orders\cite{czar2} and is
\begin{equation} 
a_{\mu}^{EW}(SM)=15.1(0.4)\times 10^{-10}
\end{equation}
The hadronic contribution from vacuum polarization corrections 
to $O(\alpha^2)$\cite{davier} and 
$O(\alpha^3)$\cite{krause} and light by light 
scattering contribution\cite{bijnens,hayakawa} together 
give\cite{davier,mr}
$a_{\mu}^{had}=673.9(6.7)\times 10^{-10}$. 
The total Standard Model contribution is then
$a_{\mu}^{SM}$=$11659159.7$ $(6.7)$ $\times 10^{-10}$.
The recent Brookhaven result gives a $2.6\sigma$ difference
between  experiment and theory\cite{brown}

\begin{equation}
a_{\mu}^{exp}-a_{\mu}^{SM}=43(16)\times 10^{-10}
\end{equation}

\section{Supersymmetric electro-weak contributions}
It is well known that the anomalous magnetic moment vanishes
in the limit of  exact supersymmetry\cite{ferrara}. 
The early analyses of $a_{\mu}$ in supersymmetric models 
with broken supersymmetry
are listed in Ref.\cite{fayet}. However, the anomalous moment
is very sensitive to the pattern of supersymmetry breaking and thus one
needs phenomenologically viable models of SUSY 
breaking for such computations.
The supergravity unified model\cite{cham,applied}
 provides such a framework and led to the modern 
analyses of supersymmetric electro-weak correction to 
$a_{\mu}$\cite{kosower,yuan}. The parameter space of the minimal
supergravity model (mSUGRA) at low energy is characterized by the
parameters $m_0,m_{\frac{1}{2}}, A_0, \tan\beta$ and sign$\mu$ 
where $m_0$ is the universal scalar mass, $m_{\frac{1}{2}}$ is the
universal gaugino mass, $A_0$ is the universal trilinear coupling,
$\tan\beta =<H_1>/<H_2>$ where, $H_2$ gives mass to the up quarks
and $H_1$ gives mass to the down quarks and leptons, and $\mu$  is
the Higgs mixing parameter which appears in the superpotential 
as $\mu H_1H_2$. 
 We reproduce here some of the results of 
Ref.\cite{yuan} where the first full one loop analysis of the supersymmetric
effect was given. The supersymmetric contribution here arises 
 from the chargino - sneutrino exchange and from the 
  neutralino - smuon  exchange so that
$a_{\mu}^{SUSY}={a_{\mu}^{\tilde W}+a_{\mu}^{\tilde Z}}$.
The chargino exchange which is typically the larger
contribution is given by\cite{chatto}
\begin{equation}
 a_{\mu}^{\tilde W}={{m^2_\mu} \over {48{\pi}^2}} {{ {A_R^{(a)}}^2} \over
{\winomass^2}}F_1(\left({{m_{\tilde \nu}} \over {m_{\tilde W_a}}}\right)^2)+
{{m_\mu} \over{8{\pi}^2}} {{A_R^{(a)}
A_L^{(a)}} \over {\winomass}} 
F_2(\left({{m_{\tilde \nu}} \over {m_{\tilde W_a}}}\right)^2)
\end{equation}
where $A_L(A_R)$ are the left(right) chiral amplitudes and are
given by\cite{chatto}  
\begin{equation}
A_R^{(1)}=-{\ecoup}\cos {\gamma_1}; \quad A_L^{(1)}={(-1)^\theta}
{{{em_\mu}\cos{\gamma_2}} \over {{2}M_W\sin\theta_W \cos {\beta}}}
\end{equation}

\begin{equation}
A_R^{(2)}=-{\ecoup}\sin {\gamma_1}; \quad A_L^{(2)}=-
{{{e m_\mu}\sin{\gamma_2}} \over
{{2}M_W \sin\theta_W\cos {\beta}}}
\end{equation}
Here $\gamma_1$ and $\gamma_2$ are the mixing angles and 
$\theta =0 (\theta =1)$  if the lighter eigenvalue of the chargino mass
matrix is negative (positive).   
The supersymmetric electro-weak corrections has several interesting
features. First under the constraint of radiative breaking of the 
electro-weak symmetry one finds that the term  proportional to 
$A_LA_R$ term in Eq.(3) dominates, and further this term itself is 
dominated by the light chargino exchange contribution. 
As is evident from Eq.(4) the 
light chargino exchange term carries with it a signature and because of 
that one finds a strong correlation between the 
sign of $\mu$ and the sign of $a_{\mu}^{SUSY}$. Thus one finds 
 that quite generally that 
  $a_{\mu}^{SUSY}>0, \mu>0$, and $a_{\mu}^{SUSY}<0, \mu<0$
  (in the standard sign convention\cite{sugra})
  except when $\tan\beta \sim 1$\cite{lopez,chatto}.
Second we also note that the dominant $A_LA_R$ term in Eq.(3) 
 has a coupling proportional to  $\sim 1/\cos\beta$.
 Because of this $a_{\mu}$ increases linearly with $\tan\beta$
  for large $\tan\beta$\cite{lopez,chatto}. 
 As will be
 discussed in Sec.3 the effects of extra dimensions on $a_{\mu}$ is 
 relatively small\cite{nyg2},  so their contribution
  to $a_{\mu}$ does not pose 
 a serious background to the supersymmetric electro-weak contribution. 
 The effect of the phases was analyzed in Ref.\cite{ing} and it is
 found that the supersymmetric contribution to $a_{\mu}$ is a 
 very sensitive function of the phases and  their  inclusion
 in the analysis can change both the sign and the magnitude of the
 supersymmetric contribution.  
 
   After the recent BNL experimental result became available\cite{brown} 
   a large number of investigations exploring
   the implications of the result for 
   supersymmetry have been 
   reported\cite{chatto2,kane,feng,gondolo,icn,marciano}. 
  In the analysis of Ref.\cite{chatto2} under the assumption of  
 CP conservation and setting 
$a_{\mu}^{SUSY}=a_{\mu}^{exp}-a_{\mu}^{SM}$, it is found 
as anticipated\cite{lopez,chatto} that the
BNL data determines the sign of $\mu$ and one finds\cite{chatto2}
$sign(\mu)=+1$.
Further, using the 2$\sigma$ error corridor of Eq.(2) one finds that 
the data implies upper bounds on the sparticle  masses.
 Thus within mSUGRA one finds\cite{chatto2}
$m_{\tilde W}\leq 650 GeV, m_{\tilde \nu}\leq 1.5 TeV$  and
$m_{1/2}\leq 800 GeV, m_0\leq 1.5 TeV$, for $\tan\beta \leq 55$. 
These results imply that the sparticles should become visible at
 the LHC and perhaps even at RUNII of the Tevatron.
 Additionally one finds that the $\mu$ sign implied by 
 the BNL  data is the sign which least restricts the supersymmetry
 parameter space under the $b\rightarrow s+\gamma$ 
 constraint\cite{bsgamma} and is also the one preferred in dark matter
 analyses. Thus the determination by the Brookhaven data that the 
 sign of 
 $\mu$ is positive is encouraging from the point of view of
 search for supersymmetric dark matter\cite{chatto2}.
  Further, as
 discussed above the CP violating phases associated with soft SUSY
 parameters can generate large contributions to $a_{\mu}$
  and affect both the magnitude
   and the sign of $a_{\mu}$\cite{ing}. 
 The above sensitivity of $a_{\mu}$  implies that 
  the BNL data, i.e., Eq.(2), can provide a strong constraint
  on the phases. This indeed turns out to be the case 
  and one finds that as much as sixty to ninety percent
   of the parameter space of the CP violating phases 
   can be eliminated by the BNL data\cite{icn}.

\section{Effect of large extra dimensions on $a_{\mu}$}
 Next we discuss the implications of large extra 
 dimensions\cite{extradim} on $a_{\mu}$(for a recent review see 
 Ref.\cite{extrareview}).
 This class of models can arise from compactifications of Type I
string theory and in models of this type the string scale
and even the fundamental Planck scale can be quite low, i.e., in the
vicinity of a few TeV\cite{extradim}. For specificity we shall 
 consider the case with one large extra dimension compactified
on $S^1/Z_2$ with compactification radius R where we assume
that the inverse radius $M_R=1/R$ is $O(TeV)$.
In this model 
  the resulting
 spectrum after compactification 
 contains massless modes with N=1 supersymmetry in 4D, which 
  precisely form the spectrum of MSSM in 4D and the
  massive Kaluza-Klein (KK) modes  form N=2 multiplets in 4D.
  The Kaluza-Klein excitations generate corrections to the 
  Fermi constant\cite{fermi} and one finds for the above model
$G_F=G_F^{SM}(1+ \frac{\pi^2}{3} \frac{M_W^2}{M_R^2})$.
With the current error corridor on $G_F^{SM}$ one finds $M_R\geq 3$TeV.
Large extra dimensions affect the value of $a_{\mu}$
from contributions via the excitations of $W,Z, \gamma$ and the
 KK correction to $a_{\mu}$ is given by\cite{nyg2}  

\begin{equation}
(\Delta a_{\mu})^{extra}= \alpha \frac{\pi}{9} 
\frac{m_{\mu}^2}{M_R^2}+
\frac{G_F m_{\mu}^2}{6\sqrt 2}
(-\frac{5}{12}+\frac{4}{3}(sin^2\theta_W-\frac{1}{4})^2) 
(\frac{M_Z^2-M_W^2}{M_R^2}) 
\end{equation}
The relative minus sign between the $M_Z^2$ and the $M_W^2$ 
terms in the last brace in Eq.(6) arises 
because the Fermi constant must be
normalized to take account of the KK correction.
Numerically the effects of the KK states is small,
i.e.,
\beq
 a_{\mu}^{extra}/a_{\mu}^{SUSY}\leq O(10^{-2})
 \eeq
Thus extra dimensions do not create a serious background to the 
SUSY contribution. The effect of extra dimensions could be enhanced 
in some models 
which, however, do not appear to be very natural\cite{desh}.
 Similar results are expected in models with strong
gravity\cite{graesser} since the fundamental Planck scale $M_*$ 
from the recent gravity experiment\cite{hoyle} is constrained so that
$M_*\geq 3.5$TeV and this scale may be as high as 50-100 TeV 
from studies of graviton emission into large extra compact
dimensions from a hot supernova core using the SN1987A data\cite{cullen}.
We note, however, that although 
 the extra dimensions are invisible in g-2 experiment
their effects could still become visible at accelerators with large enough
energies\cite{extrareview,nyy,peskin}. Additional analyses within
the framework of extra dimensions can be found in Ref.\cite{g2extra}.

\section{ Conclusion}
SUSY provides the most natural explanation of the difference 
$a_{\mu}^{exp}-a_{\mu}^{SM}$ seen at BNL.  
The effect was already predicted within the framework of SUGRA models
where it was known since the early to mid nineteen eightees 
that the supersymmetric correction could be as large or 
larger than the Standard Model electro-weak correction\cite{yuan}.
 A detailed analysis of the implications of the BNL experiment in 
 mSUGRA shows that if the size of the new physics  effect seen 
 by Brookhaven persists it would 
imply the existence of sparticles accessible at the LHC. Further,
if SUSY is the right explanation of the difference seen in the
BNL experiment, then the existence of a Higgs field as a fundamental 
field (as opposed to a composite field) is implied. In SUGRA
unified models there is an upper limit of about 130 GeV
 for the lightest neutral Higgs boson
 within the usual naturalness limits on sparticle masses\cite{ccn}
 and thus one expects this Higgs boson
to become visible at RUNII of the Tevatron with appropriate 
integrated luminosity. Additionally the possibility of finding
a sparticle at RUNII is not excluded. 
It is also found that the BNL data imposes impressive constraints
 on the phases of soft SUSY breaking parameters eliminating a large
  part of the parameter space  of these phases.
 The effect of extra space-time dimensions on $a_{\mu}$ was also
 discussed and it is concluded that extra dimensions  do not generate
 an effect comparable to the supersymmetric electro-weak effect.
There are other possibilities 
not discussed here such as of a light higgs\cite{haber},
 lepto-quarks, composite models, techni-color
and extra gauge bosons as possible sources for a large correction
to $a_{\mu}$ (for a general review of these see 
Czarnecki and Marciano in Ref.\cite{marciano} and for a
model independent analysis see Ref.\cite{einhorn}). However,
 of all the scenarios mentioned above the possibility that the 
observed effect is arising from supersymmetry appears to us to be
 the most compelling.

In the coming months additional data collected in runs in the 
year 2000 will be analyzed and one expects that the experimental
error will reduce further. The central issue of course is the size 
of the difference $a_{\mu}^{exp}-a_{\mu}^{SM}$ and the associated
 error corridor. Here the crucial question concerns the size of the
hadronic correction and the error associated with it. Since much
of the error arises in the energy domain of less than 2 GeV in
the $e^+e^-\rightarrow hadrons$ cross section more accurate data in this
region will certainly help reduce the errors\cite{mr}. Another sensitive 
issue is the light by light contribution to the hadronic 
error. Although a computation of this correction relies entirely
on theoretical models, it is comforting that 
two independent analyses\cite{bijnens,hayakawa} are in 
agreement on the overall sign and  also in fair agreement on
the magnitude of this contribution.
These issues are expected to be explored in greater depth in
the coming months and along with the expected more accurate experimental 
measurement of $a_{\mu}$ at 
 Brookhaven, the revised version of Eq.(2) will pin down the scale
of weak scale supersymmetry even more precisely.

\section{Acknowledgements}
The work reported here is a review of the collaborative work with 
Richard Arnowitt, Ali Chamseddine, and T.C. Yuan in the early
nineteen eightees when the modern analysis of g-2 was carried out, 
of the more recent works with Utpal 
Chattopadhyay and Tarek Ibrahim, and of the work with Masahiro Yamaguchi
on the effect of large extra dimensions on g-2. 
This research was supported in part by the NSF grant PHY-9901057.
\end{document}